**Statistical Study and Live Catalogue of Multi-Spacecraft $^3$He-Rich Time Periods over Solar Cycles 23, 24, and 25**

**Short Title: Live Catalogue of $^3$He-Rich Time Periods**

S.T. Hart[1,2]*, M.A. Dayeh[2,1], R. Bučík[2], M.I. Desai[2,1], R.W. Ebert[2,1], G.C. Ho[3], G. Li[4], G.M. Mason[3]

[1]The University of Texas at San Antonio, San Antonio, TX, USA, 78249
[2]Southwest Research Institute, San Antonio, TX, USA, 78238
[3]Johns Hopkins University Applied Physics Laboratory, Laurel, MD, USA, 20723
[4]University of Alabama in Huntsville, Huntsville, AL, USA, 35899
*Corresponding author: samuel.hart@contractor.swri.org

**Abstract**

Using ion measurements from Ultra-Low-Energy Isotope Spectrometer (ULEIS) observations onboard *Advanced Composition Explorer* (ACE) and Solar Isotope Spectrometer (SIS) observations onboard the *Solar Terrestrial Observatory* (STEREO)-A and STEREO-B spacecraft, we have identified 854 $^3$He-rich time periods between 1997 September and 2021 March. We include all event types with observed $^3$He enhancements such as corotating interaction regions (CIRs), gradual solar energetic particle (SEP) events, interplanetary shocks, and impulsive SEP events. We employ two different mass separation techniques to obtain $^3$He, $^4$He, Fe, and O fluences for each event, and we determine the $^3$He/$^4$He and Fe/O abundance ratios between 0.32 – 0.45 MeV/nucleon and 0.64 – 1.28 MeV/nucleon. We find a clear correlation in the $^3$He/$^4$He and Fe/O abundance ratios between both energy ranges. We find two distinct trends in the $^3$He/$^4$He vs. Fe/O relation. For low $^3$He/$^4$He values, there is a positive linear correlation between $^3$He/$^4$He and Fe/O. However, at $^3$He/$^4$He ~ 0.3, Fe/O appears to reach a limit and the correlation weakens significantly. We provide a live catalogue of $^3$He-rich time periods that includes the robust determination of the onset and end times of the $^3$He enhancements in SEP-associated periods for different types of events observed my multiple spacecraft. This catalogue is available for public use. New releases will follow after major additions such as adding new periods from new missions (e.g., Parker Solar Probe and Solar Orbiter), identifying event types (impulsive SEP events, etc.), or adding new parameters such as remote observations detailing characteristics of the events' active regions.

**1. Introduction**

The Sun is capable of accelerating particles to >MeV energies. Coronal mass ejections (CMEs) and corotating interaction regions (CIRs) can drive interplanetary (IP) shocks or large compression regions that accelerate particles via diffusive shock acceleration (DSA, e.g. Decker 1981, Lee 1983, Zank et al. 2015, Giacalone et al. 2002). Solar flares may also accelerate particles via magnetic reconnection within solar active regions by converting magnetic energy into kinetic energy that accelerates and heats particles within the reconnection site (e.g. Wang et al. 2006; Bučík 2020). The resulting energetic particle populations from these processes are referred to as solar energetic particles (SEPs). The primary candidate for accelerating energetic particles at IP shocks is the shock-drift mechanism at quasiperpendicular shocks (Decker, 1981) and first-order Fermi acceleration mechanism at quasi-parallel shocks (Lee, 1983). The strongest shockwaves originate at the front of fast CMEs, producing gradual SEPs (GSEPs) that often increase in intensity over several days until peaking at shock passage before decreasing quickly in intensity. DSA is a highly efficient acceleration method and thus tends to produce SEPs with energies greater than tens of MeV/nucleon. (Desai & Giacalone et al. 2016). Additionally, IP shocks can propagate over a wide range of heliospheric longitudes. As a result, GSEP events observed near Earth originate from a broad range of source locations on the solar disc (e.g. Reames 1999).

Magnetic reconnection occurs as a result of compressing anti-parallel components of magnetic field lines (Gonzalez & Parker 2016). Particles at reconnection sites are stochastically accelerated by significant amounts of magnetic turbulence (Petrosian et al. 2012). In some reconnection events, the newly reconnected field lines are open (insofar as they close well beyond the orbit of Earth) and the energetic particles along these field lines are injected into IP space and observed as EUV jets (Bučík 2020). This process is often referred to as interchange reconnection, and the escaping particles are called impulsive SEPs (ISEPs) due to their short-lived nature. Unlike GSEPs, ISEPs are rarely observed to have energies greater than 10 MeV/nucleon. In fact, GSEP event identification criteria often include a significant abundance enhancement in SEPs greater than 10 MeV/nucleon to exclude ISEP events (e.g., Cane et al. 2010, Desai et al. 2016, the NOAA proton event list).

The difference in elemental composition of GSEPs and ISEPs can be attributed to some combination of their source region as well as their acceleration mechanisms (Reames & Ng 2004). GSEPs, produced in the upper corona and in interplanetary space, have similar compositions to the solar wind or corona (e.g., Desai et al. 2006), whereas most observed ISEPs occur lower corona and chromosphere and contain enhanced Fe/O abundance ratios as well as $^3$He/$^4$He abundance ratios of unity (e.g., Mason et al. 2007). For reference, the $^3$He/$^4$He abundance ratio is less than 1 in $10^{-4}$ in our solar system (Gloeckler & Geiss 1998) making the $^3$He enhancement in ISEPs one of the largest known fractionation processes and bringing rise to their alternate name, $^3$He-rich SEPs.

$^3$He-rich SEP events are of particular interest because of their peculiar behavior. First, the nature behind enhanced $^3$He/$^4$He abundance ratios is still unresolved. Theoretical studies attribute the $^3$He enhancement to wave-particle interactions in the flaring regions (Temerin & Roth 1992, Liu et al. 2004, Zhang 1995, see Klecker et al. 2006 and Miller 1998 for review). Second, although $^3$He-rich SEP events are produced in solar flares, the $^3$He/$^4$He abundance ratio and $^3$He fluences do not correlate with the soft X-ray peak flux of solar flares (Nitta et al. 2006). They are more closely correlated with type III radio bursts (emissions generated by escaping non-relativistic electrons at the reconnection site), but a recent study by *Köberle et al.* (2021) shows that not all $^3$He-rich SEP events have an associated electron enhancement. Third, $^3$He-rich SEP events display considerable interevent variations in relative abundances and spectral shape (e.g., Reames et al. 1994, Mason et al. 2007), suggesting that a single acceleration mechanism is insufficient and instead supports a combination of multiple independent acceleration mechanisms. Fourth, $^3$He at suprathermal energies is commonly observed to have enhanced abundances in IP space during quiet-times throughout the solar cycle (Dayeh et al. 2009, Wiedenbeck et al. 2005), suggesting that interchange reconnection may happen on small scales near active regions, similar to the theory of nanoflares (Parker et al. 1988). Finally, $^3$He-rich SEP events were originally believed to have access to only a narrow range of pre-existing interplanetary magnetic field lines, and they were observed to have source regions on a tight longitudinal band on the western hemisphere of the Sun (e.g., Reames 1999). However, more recent studies (Nitta et al. 2015 and references therein) show much broader longitudinal distributions. Furthermore, observations of $^3$He-rich SEP events with longitudinally separated spacecraft show that a single $^3$He-rich event can be distributed 180° in heliospheric longitude (Wiedenbeck et al. 2010, 2013). Following these observations, simulations have been developed seeking to explain this phenomena. Magnetic field models near the solar surface show that magnetic field lines near the open-closed boundary can be widely dispersed at the source surface, resulting in the occasional broad longitudinal dispersion of $^3$He-rich SEP events (Scott et al. 2018). Once in the heliosphere, magnetic field lines can wander, deviating significantly from the ideal Parker spiral and altering the path of flowing energetic particles (Howes & Bourouaine 2017, Moradi & Li 2019, Bian & Li 2021, 2022). Still, the dominant process governing the longitudinal

distribution of $^3$He-rich SEP events is not fully understood, and remains an active topic of debate within the community.

In this work, we identify and examine the properties of $^3$He-rich time periods observed by the *Advanced Composition Explorer* (ACE; 1997 Sept. – 2021 Mar.) and the *Solar Terrestrial Observatory's* Ahead (STEREO-A; 2007 – 2021 Mar.) and Behind (STEREO-B; 2007 – 2014) spacecraft. In total, we have identified 854 events at the time of publication. We use a simple mass cutoff for measurements on instruments with high mass resolution (ACE/ULEIS), and a triple-peaked Gaussian on top of a linear background for instruments with lower mass resolution (STEREO-A & B/SIT) to extract the fluences of $^3$He, $^4$He, Fe, and O. We then determine the $^3$He/$^4$He and Fe/O abundance ratio at 0.32 – 0.45 MeV/nucleon and 0.64 – 1.28 MeV/nucleon, and we comment on their distributions and relations.

Ultimately, we introduce a new live catalogue of $^3$He-rich time periods. These time periods include all possible event types such as CIRs, IP shock passages, GSEP events, and ISEP events. The catalogue includes all of the time periods identified in this paper, and we will continue to update it as current missions observe more events during solar cycle 25 and beyond. The intent is to also include events from Parker Solar Probe and Solar Orbiter. In addition to event start/stop dates and abundance ratios, the future of this catalogue will include all subsequent data relevant to the event such as the event type (e.g., GSEP, ISEP, etc.), parameters inferred from *in-situ* observations (e.g., fluences, travel path length, etc.) and characteristics of the source region (e.g., X-ray flare class, EUV flare shape, type III radio burst), flag for the multi-spacecraft detection, and potentially relevant IP space parameters in the inner heliosphere (e.g., ambient times, preceding CMEs, shocks, etc.). This catalogue is live and publicly available and will continue to build into a unique and comprehensive catalogue to be used in the process of answering fundamental questions pertaining to the nature of $^3$He-rich SEP events noted above.

**2. Data & Instrumentation**

We use the Ultra-Low-Energy Ion Spectrometer (ULEIS; Mason et al. 1998) onboard ACE (Stone et al. 1998) and the Suprathermal Ion Telescope (SIT; Mason et al. 2008) onboard both STEREO A & B (Kaiser et al. 2008). Both ULEIS and SIT are mass spectrometers equipped with time-of-flight sensors to allow for mass resolution between 0 – 80 amu in the 0.04 MeV/nucleon to ~5 MeV/nucleon kinetic energy range. ULEIS data has a 24-second time cadence and STEREO data has a one-minute time cadence. In both cases, we study the 0.16 MeV/nucleon to ~5 MeV/nucleon energy range with masses between 0 – 80 amu.

## 3. Methodology

### *3.1 Selecting $^3$He-Rich Time Periods*

$^3$He-rich time periods are determined visually using the publicly available (Level 3), 4-day mass spectrogram plots provided by the ACE Science Center (izw1.caltech.edu/ACE) and the STEREO Science Center (izw1.caltech.edu/STEREO). We take several factors into account when selecting the time range of each event. For each event, we require an increase in the count rates near 3 amu ($^3$He) in the mass spectrograms. We exclude events whose $^3$He enhancement is solely due to H and $^4$He spillover into the $^3$He mass range. An example of our selection process is illustrated in Figure 1 (see Figure 2 for STEREO). Figure 1 is a modified version of the publicly available 4-day browse plots, and it shows an ISEP event observed by ACE beginning on 1999 Sept. 30 and lasting four days. Figure 1b shows a 2 – 80 amu mass spectrogram between 0.32 – 3.6 MeV/nucleon. The dashed vertical line indicates the start of the count rate increase near 3 amu, signifying a $^3$He enhancement. Figure 1a shows the flux time profiles of $^4$He, O, & Fe between 0.16 – 0.226 MeV/nucleon, noting that the Fe (orange) & O (red) flux temporal profiles nearly overlap for the duration of the event, whereas other heliospheric populations have O fluxes typically a factor of 10 greater than that of Fe (Desai et al. 2006). Figure 1c displays the one-over-ion speed plot along with the red line being the approximate velocity dispersion of this particular event. We repeat this process for ACE and both STEREO spacecraft to obtain 854 $^3$He-rich time periods since September 1997.

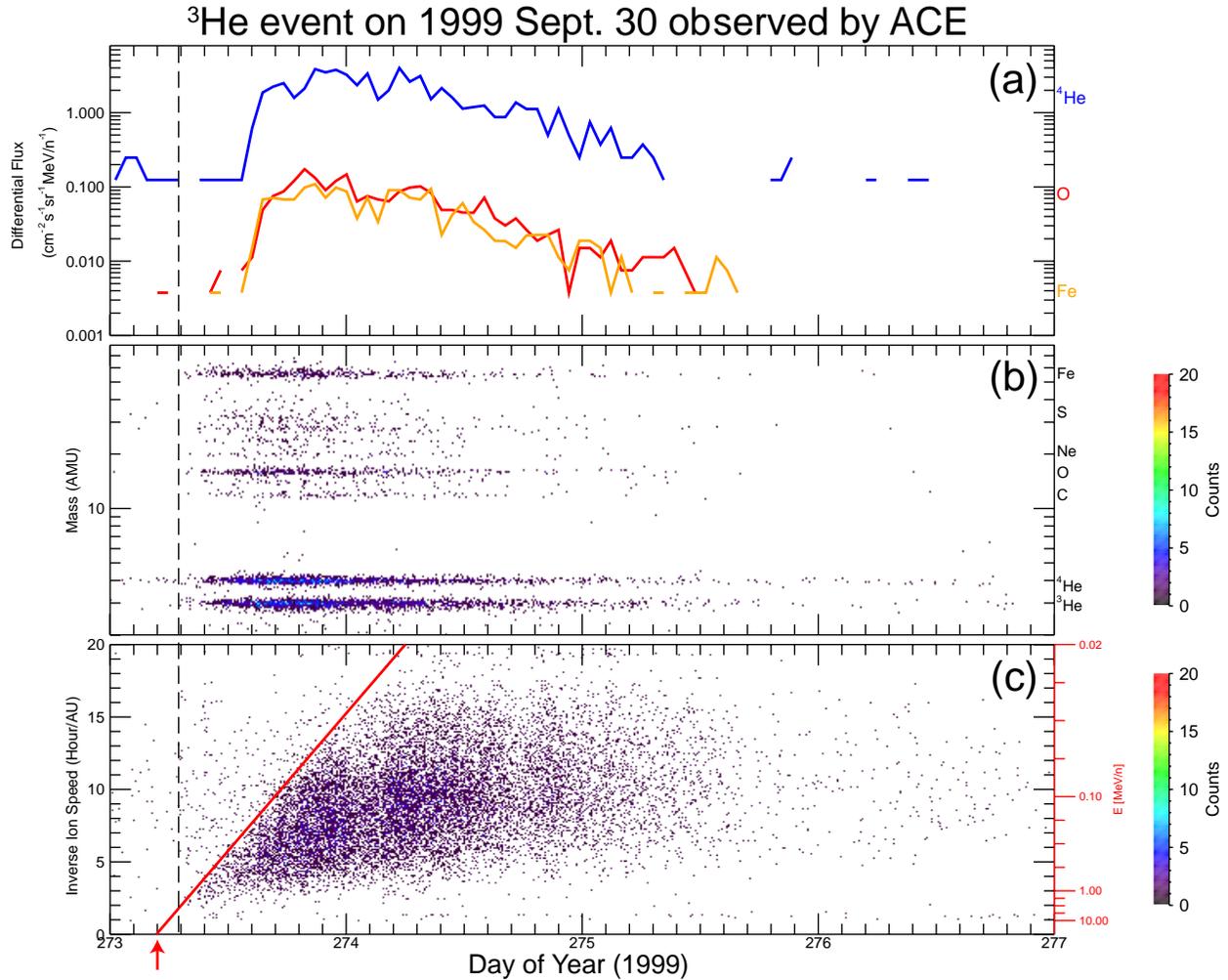

**Figure 1.** (a) Hourly flux values of $^4$He, O, and Fe at 0.19 MeV/nucleon; clearly shows an overlapping flux temporal profile of Fe & O commonly observed in $^3$He-rich SEP events. (b) Mass spectrogram between 2 – 80 amu with an energy range of 0.32 MeV/nucleon – 3.6 MeV/nucleon. The most common species are labelled on the right side on the plot. Starting at DOY 273.3 (dashed vertical line), the mass spectrogram shows an enhancement of all of the most common species, most notably $^3$He, signaling an observation of a $^3$He-rich SEP event that lasts for a few days. (c) One-over-ion speed plot of masses 10 – 80 amu. The red axis on the right side of the plot shows the energy-per-nucleon that corresponds to the inverse speed. The slanted red line shows the approximate arrival times of particles with different energies. The dispersion's intersection with the time axis yields injection time (red arrow), and the slope reveals path length of the ion travel to 1 AU.

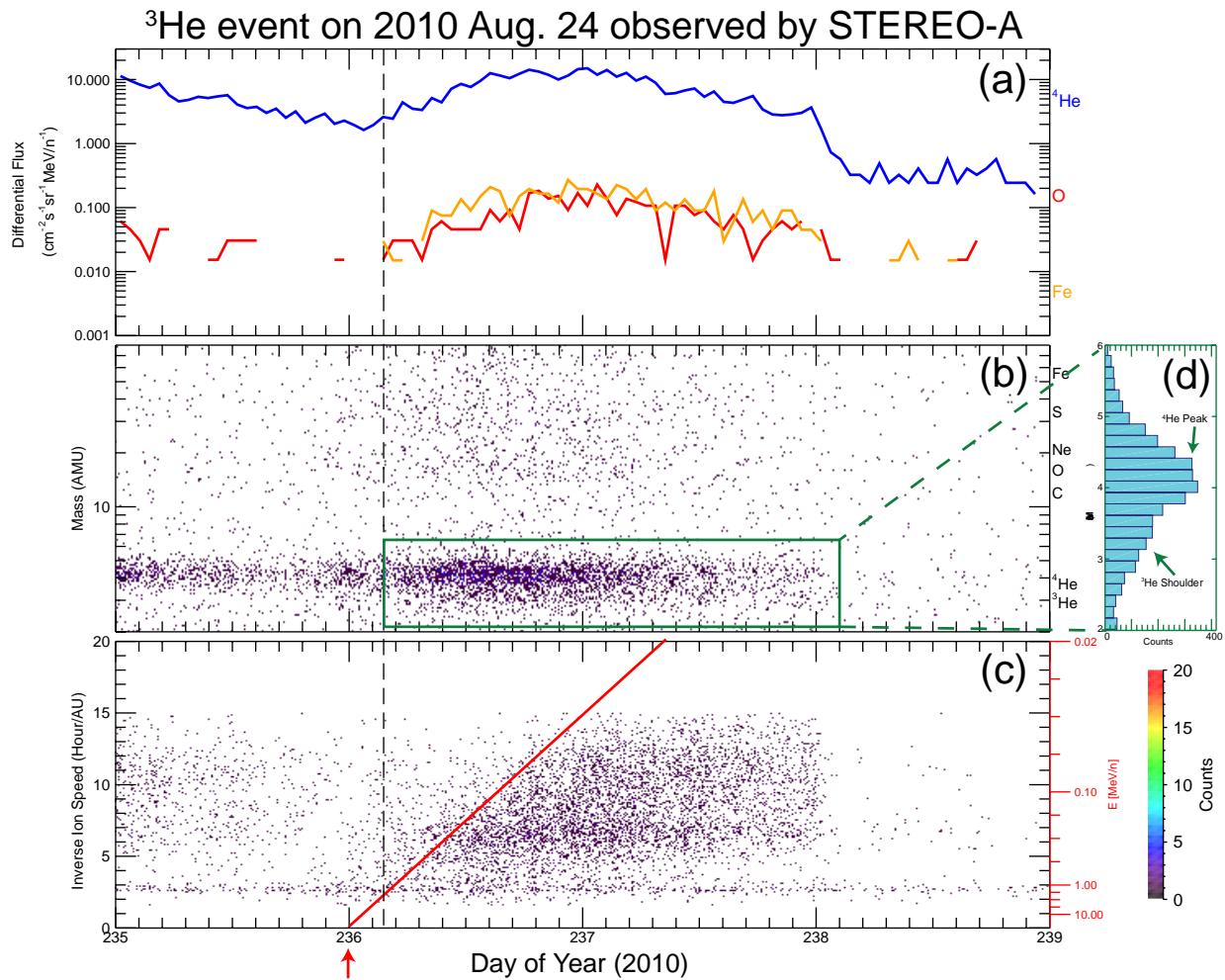

**Figure 2.** Same format as Figure 1, but for an ISEP event observed by STEREO-A at the start of 2010 August 24. Note again that the Fe & O flux temporal profiles overlap in (a). Comparing (b) to Figure 1b, it is clear that the mass resolution of STEREO/SIT is broader than that of ACE/ULEIS. To help clarify the ³He enhancement, we include the time-integrated mass histogram (d) of the contents within the green box to highlight the ³He "shoulder" next to the ⁴He peak. For more detailed images, see §3.2.2.

### 3.2 Determining Abundance Ratios

#### 3.2.1 ³He/⁴He using ACE/ULEIS

The mass resolution of ULEIS onboard ACE is the narrowest of the surveyed instruments in this catalogue to date, with a $\sigma_m < 0.15$ amu (Mason et al. 1998). The 1 amu mass difference of ³He and ⁴He means the isotopic overlap begins over three standard deviations out from the mass center, i.e., these isotopes are nearly completely separable and any overlap is negligible. As a result, we follow a similar mass cutoff approach as *Desai et al.* (2001) splitting ³He and ⁴He at 3.3 amu to determine the relative abundances. We accumulate observations over the duration of the enrichment period and bin the observations into a 1-D mass histogram for two distinct energy ranges (0.32 – 0.45 MeV/nucleon and 0.64 – 1.28 MeV/nucleon). Abundance ratios and errors are propagated with the following statistics:

$$R = \frac{N(3He) * Q(4He)}{N(4He) * Q(3He)} \tag{1}$$

$$\sigma_R = R \sqrt{\frac{\sigma(3He)^2}{N(3He)^2} + \frac{\sigma(4He)^2}{N(4He)^2}} \tag{2}$$

where $R$ and $\sigma_R$ are the abundance ratio and its propagated uncertainty, respectively, $N(X)$ and $\sigma(X)$ are the total number and uncertainty of ³He or ⁴He counts during the enhancement for a given energy range, and $Q(X)$ is the detector efficiency of $X$ species in said energy range. When $N(X)$ is obtained by simple counting, we apply Poisson statistics and equation (2) reduces to:

$$\sigma_R = R \sqrt{\frac{1}{N(3He)} + \frac{1}{N(4He)}} \tag{3}$$

The mass histograms of the September 30, 1999 event are shown in Figure 3, where the observations are accumulated from 1999 September 30 07:00:00 UTC through 1999 October 4 00:00:00 UTC. The vertical line in each plot indicates the mass cutoff, and using equations (1) & (3) we obtain a ³He/⁴He abundance ratio of 0.94 ± 0.04 for the low energy range and 2.02 ± 0.18 for the high energy range.

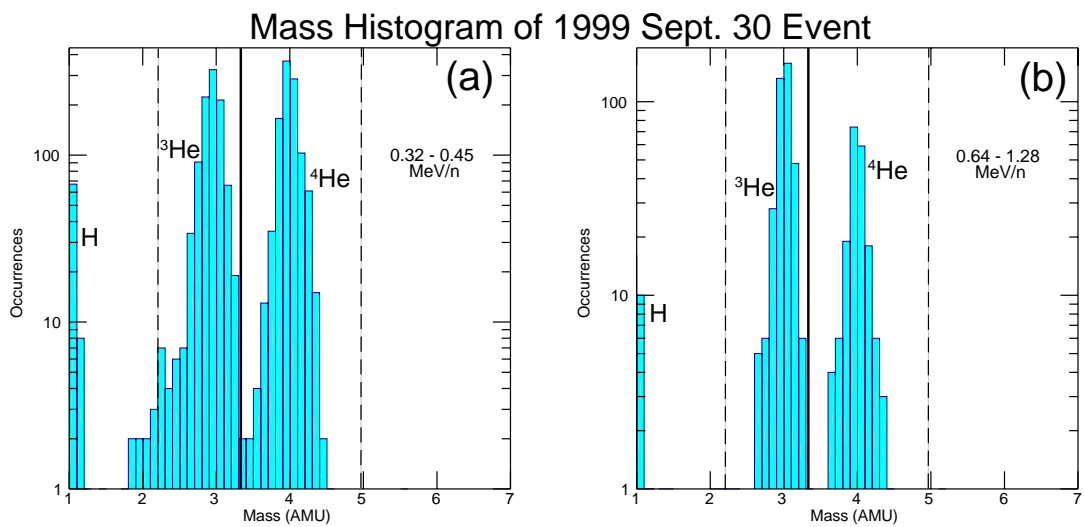

**Figure 3.** Mass histograms highlighting the separation of ³He & ⁴He for the ³He-rich SEP event that occurred on 1999 Sept. 30 observed by ACE. The black line

shows the separation mass of 3.3 amu. Anything greater is considered $^4$He up to 5 amu; less is considered $^3$He down to 2.2 amu (vertical dashed lines). We measure $^3$He/$^4$He abundance ratios of 0.94 ± 0.04 (a) for 0.32 – 0.45 MeV/nucleon and 2.02 ± 0.18 (b) for 0.64 – 1.28 MeV/nucleon.

*3.2.2 $^3$He/$^4$He using STEREO/SIT*

The mass resolution of SIT onboard both STEREO spacecraft is inferior to that of ULEIS, having a $\sigma_m/m = 0.1$ (Mason et al. 2008) resulting in $\sigma_{3He} = 0.3$ amu and $\sigma_{4He} = 0.4$ amu. Thus, the two species overlap just outside one standard deviation from each other and the stringent mass cutoff technique used with ULEIS will not suffice for SIT. However, the mass distribution of an individual species observed by SIT can be well approximated by a Gaussian. We assume, then, that the combined mass distribution of H, $^3$He, and $^4$He is a resolvable triple-peaked Gaussian, and the abundance ratio of $^3$He and $^4$He is the ratio of their integrated Gaussian distributions.

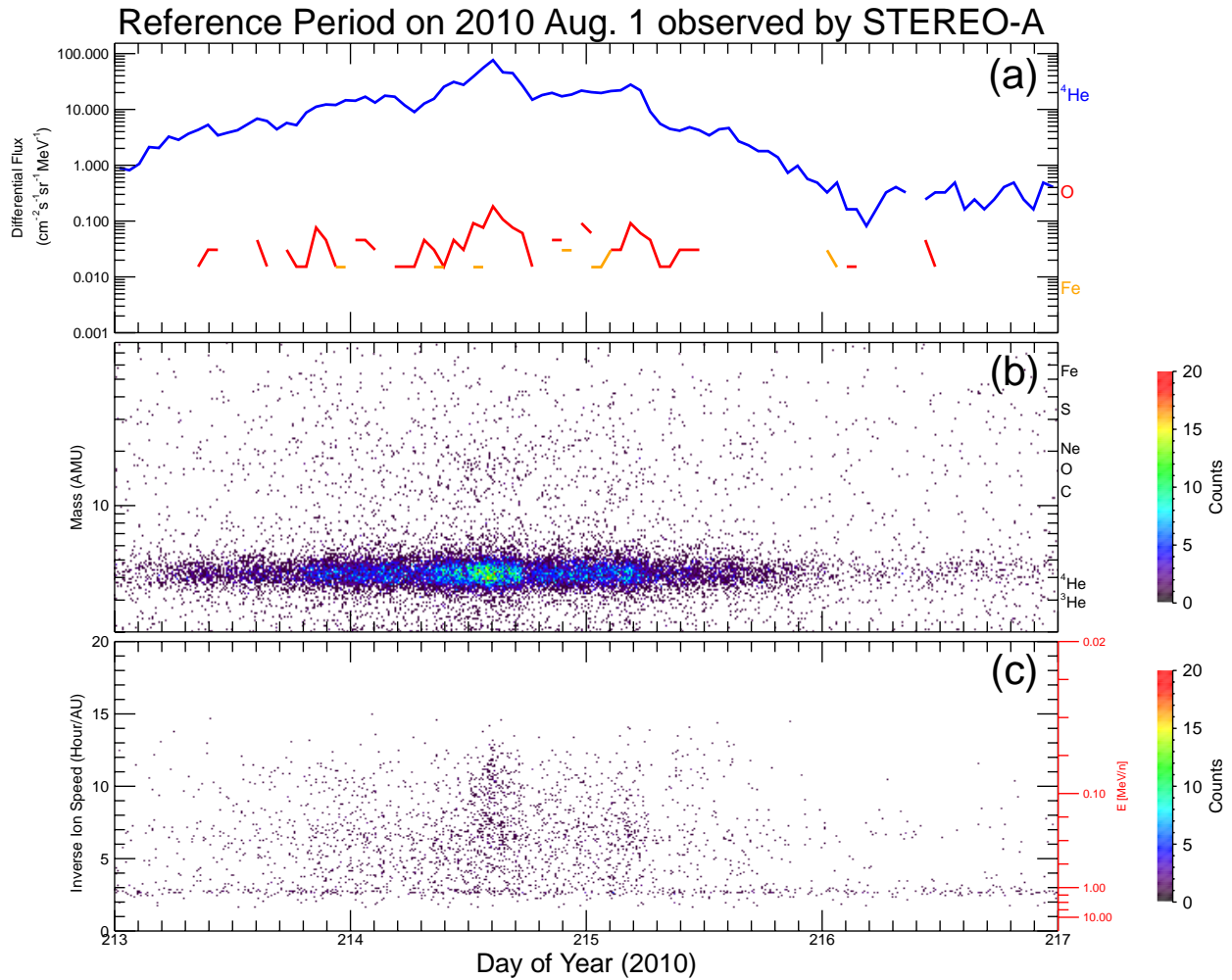

**Figure 4**. Four-day-plot of reference period observed by STEREO-A. There are no signatures of a $^3$He enhancement, evident in (b). We use the mass distributions of reference events to resolve the $^3$He fluences of $^3$He-rich time periods.

To better determine the initial estimates used in the Gaussian fitting, we make use of "reference periods" to obtain estimates on all the parameters. We define a "reference period" to be an ion enhancement observed by SIT with no clear indicators of $^3$He enhancement that occurs within approximately six (6) months of a $^3$He-rich time period of interest. The six month separation limit accounts for general instrument response changes over the lifetime of the mission. Figure 4 shows a 4-day plot of one such reference period observed between 2010 August 1 & August 5. Figure 5 shows the mass histogram of the reference period, noting again that *there is no $^3$He enhancement during this time period*. We fit a double-peaked Gaussian distribution above a linear background to the mass histogram of the reference period, and we use the best-fitted parameters of the H and $^4$He mass distributions as initial estimates for the $^3$He-rich events. The means and standard deviations of each species' distributions are listed in the same color as their respective curves. The need for the reference periods becomes evident when noticing the ~[2, 6]% decrease in the mean mass of [H, $^4$He], moving from [0.96, 4.23] amu at the lower energy bin (Figure 5a) down to [0.94, 3.99] amu at the higher energy bin (Figure 5b). Reference periods allow us to account for energy dependent mass drifting and mass resolution.

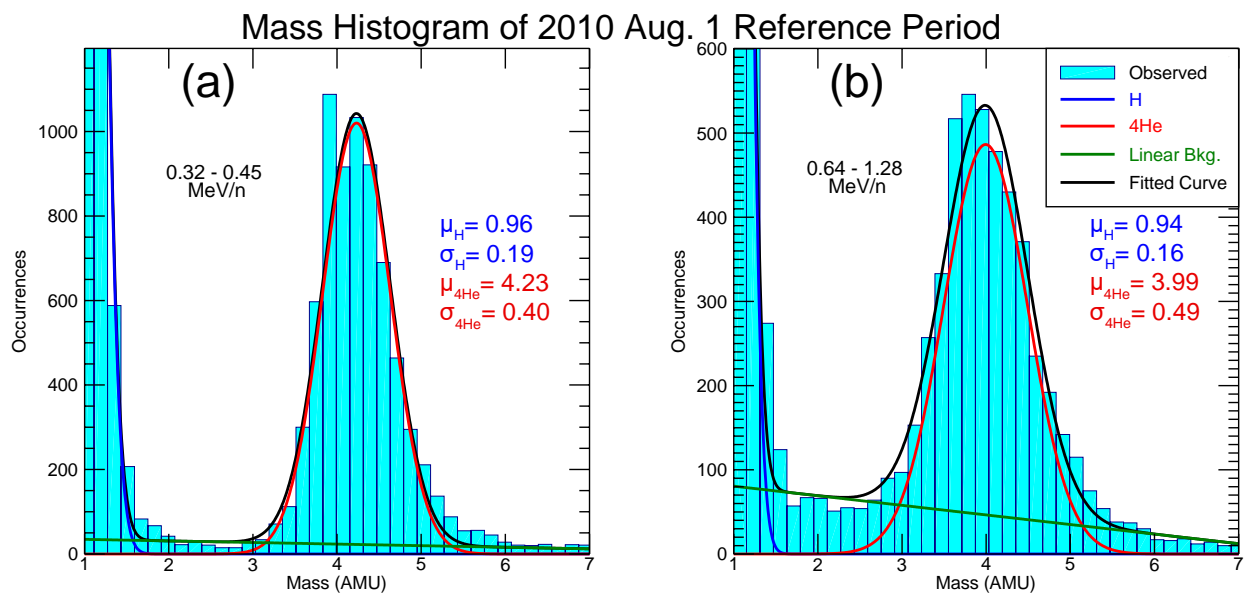

**Figure 5.** Mass histogram of 1 – 7 amu between 0.32 – 0.45 MeV/nucleon (a) and 0.64 – 1.28 MeV/nucleon (b) of a reference period observed by STEREO-A. This reference period has significant enhancement in $^4$He and H, but it does not show any $^3$He enhancement. We take the mean and standard deviation of the best-fitted gaussians (shown in blue and red, matching their respective curve) and use them as approximate parameters when separating masses for $^3$He-rich time periods.

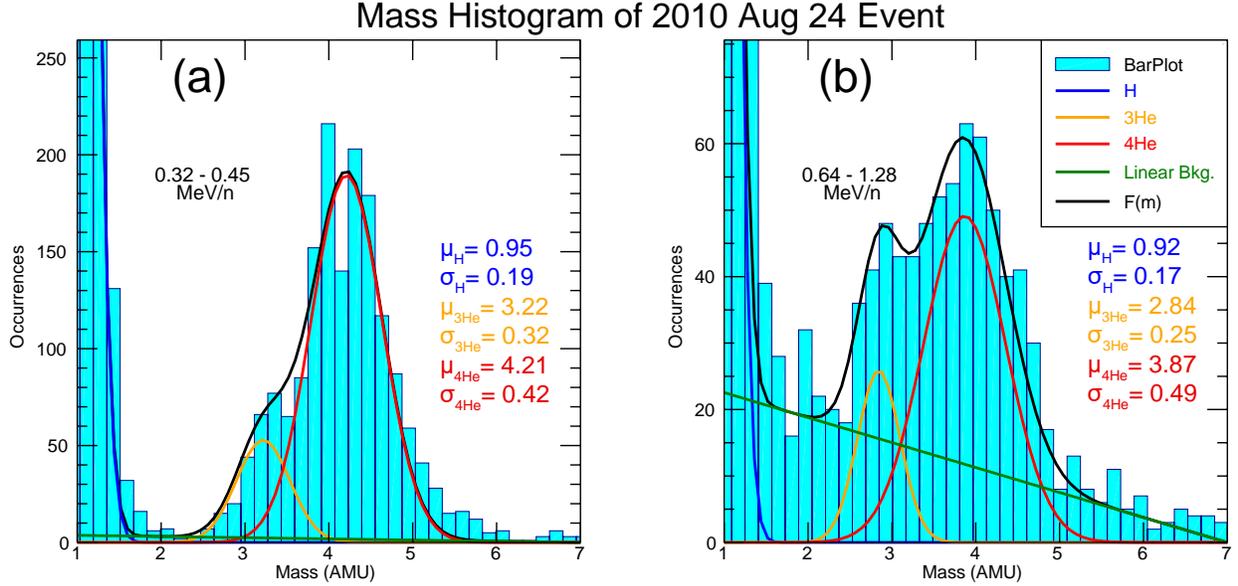

**Figure 6.** Mass histograms of $^3$He event (see Figure 2) observed by STEREO-A with $^3$He & $^4$He separated. The mean and standard deviation of H, $^3$He, and $^4$He are listed in the same color as their respective curves. Integrating the best-fitted gaussian, normalizing to a standard bin size, and accounting for the detector efficiency yields an abundance ratio of be 0.25 ± 0.02 between 0.32 – 0.45 MeV/nucleon (a). The same method yields an abundance ratio of 0.37 ± 0.04 between 0.64 – 1.28 MeV/nucleon (b).

To extract $^3$He observed by STEREO, we fit a triple-peaked Gaussian on a linear background between 0 – 8 amu with the form:

$$F(m) = A(m - m_o) + \sum_X C_{o,X} e^{-\frac{(m-\mu_X)^2}{2\sigma_X^2}} \qquad (4)$$

where $F(m)$ is the best-fitted curve, $A$ and $m_o$ are parameters contributing to the linear background, $X$ is iterated over H, $^3$He, and $^4$He, and $C_o$, $\mu$, and $\sigma$ are the height, mean, and standard deviation of each species' best-fitted Gaussian distribution. We use Poisson uncertainties for each histogram bin. Our process is shown in Figure 6 using a $^3$He-rich SEP event that began on 2010 Aug. 24

04:00:00 UTC and concluded on 2010 Aug. 26 02:00:00 UTC (see Figure 2). To produce the best-fitted curve, we input the mass mean and standard deviation values of H and $^4$He from the reference period shown in Figure 5 as initial estimates. We use IDL's MPFIT package (Markwardt 2009) to impose strict boundary limitations on the reference period parameters, and we impose looser boundary limitations on the $^3$He parameters based on the knowledge that it should behave similarly to $^4$He with a mean mass near 3 amu and a width of approximately 0.3 amu. The resulting best-fitted curve is shown in black. The individual expressions are shown in color with the addition of $^3$He. In both histograms, there is a clear enhancement near 3 amu compared to its reference period (Figure 5). We approximate the number of counts of each species to be the integral of its best-fitted Gaussian curve divided by the 0.16 amu bin size. Mathematically, this is expressed as:

$$N(X) \approx \frac{1}{B}\int_{-\infty}^{\infty} C_{o,X} e^{-\frac{(m-\mu_X)^2}{2\sigma_X^2}} \quad (5)$$

$$\sigma(X) \approx N(X) - \frac{1}{B}\int_{-\infty}^{\infty} \left(C_{o,X} - \sqrt{VAR(C_{o,X})}\right) e^{-\frac{(m-\mu_X)^2}{2\sigma_X^2}} \quad (6)$$

where $N(X)$ and $\sigma(X)$ are the approximate number of counts and uncertainty, respectively, of $^3$He or $^4$He and $B$ is the bin size of the histogram. $VAR(C_{o,X})$ is the variance of the height of the best-fitted Gaussian, $C_{o,X}$, obtained from the derived covariance matrix of the Gaussian fit. Effectively, the uncertainty in counts is the difference in area of the best-fitted Gaussian integral and the same Gaussian integral but whose height is reduced by $VAR(C_{o,X})^{1/2}$. Plugging equation (5) into (6) yields:

$$\sigma(X) \approx \frac{1}{B}\int_{-\infty}^{\infty} \sqrt{VAR(C_{o,X})} e^{-\frac{(m-\mu_X)^2}{2\sigma_X^2}} \quad (7)$$

Applying equations (5) & (7) to the 2010 August 24 event, we obtain 260 ± 26 $^3$He counts and 1237 ± 44 $^4$He counts between 0.32 – 0.45 MeV/nucleon, and we obtain 101 ± 19 counts of $^3$He and 363 ± 28 counts of $^4$He between 0.64 – 1.28 MeV/nucleon. At 0.32 – 0.45 MeV/nucleon, the $^3$He and $^4$He efficiencies are 0.245 and 0.287, respectively. At 0.64 – 1.28 MeV/nucleon, the efficiencies are 0.171 and 0.229. Using equations (1) & (2), we calculate the $^3$He/$^4$He abundance ratios to be 0.25 ± 0.03 and 0.37 ± 0.08, respectively.

*3.2.3 Fe/O using ACE and STEREO*

In addition to $^3$He/$^4$He, we also measure Fe/O in the same energy ranges. Figure 7 shows heavy ion mass histograms (10 – 100 amu) of the two events detailed above in the 0.32 – 0.45 MeV/nucleon energy range. For both spacecraft, we consider Fe and O observations to have mass between 45 – 80 amu and 14 – 19 amu, respectively. Accounting for efficiencies, the observed Fe/O abundance ratios for the 1999 Sept. 30 event observed by ACE is 0.87 ± 0.08 (Figure 7a) and 1.32 ± 0.20 for the 2010 Aug. 24 event observed by STEREO-A (Figure 7b).

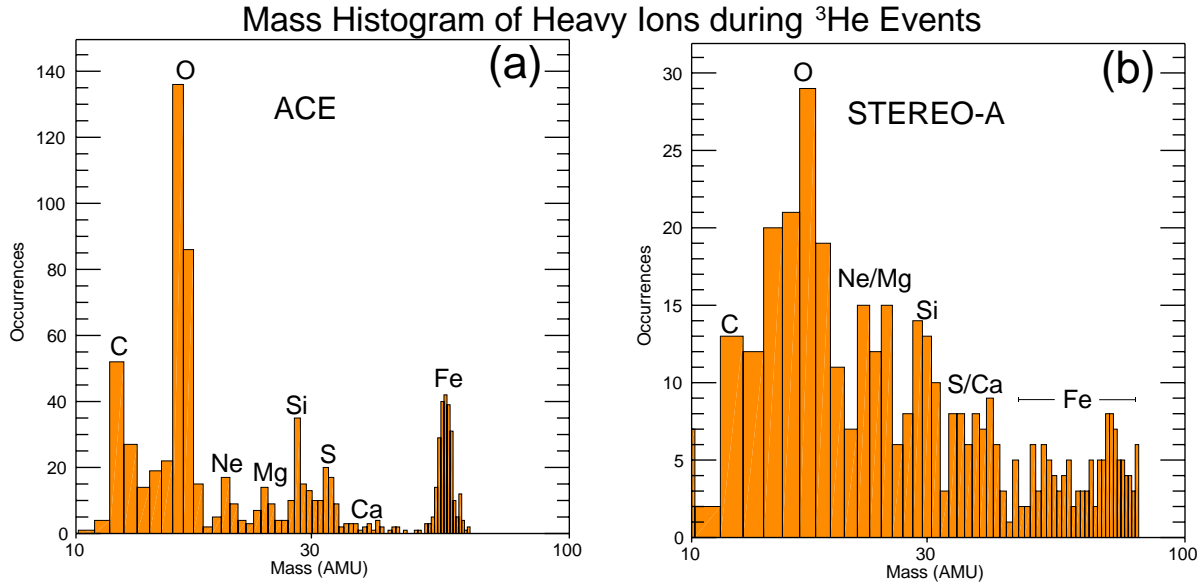

**Figure 7.** Heavy ion abundances during the 1999 Sept. 30 event observed by ACE (a) and 2010 Aug. 24 observed by STEREO-A (b) between 0.32 – 0.45 MeV/nucleon. We measure the Fe and O abundances by summing incident ion observations with masses between 45 – 80 amu and 14 – 19 amu, respectively.

## 4. Results & Discussion

Using observations from ACE, STEREO-A, and STEREO-B, we have identified 854 $^3$He-rich time periods between 1997 September – 2021 March. For each event, we use the species separating methods defined in §3 to determine both the $^3$He/$^4$He and Fe/O abundance ratios for two distinct energy ranges. The distributions and correlations of the abundance ratios are shown in Figures 8 – 10. For each figure, we only show time periods where both sets of abundance ratios have a fractional uncertainty less than 50%, i.e., periods with larger fluences.

Figure 8 shows the $^3$He/$^4$He abundance ratios at high energies (0.64 – 1.28 MeV/nucleon) vs. the $^3$He/$^4$He abundance ratios at low energies (0.32 – 0.45 MeV/nucleon) along with the line of symmetry. The 1-D distributions of the relative abundances for each spacecraft are shown along their respective axis, and the mean, median, and standard deviation of the distributions are shown in a table in the upper right corner. The median $^3$He/$^4$He abundance ratio is greater than 0.1 for all spacecraft at both energies. Additionally, the $^3$He/$^4$He abundance ratio at high energies is consistently larger than at low energies, consistent with previous studies (e.g. Mason et al. 2007, Nitta et al. 2015, Bučík et al. 2018a, Mason et al. 2021, Bučík et al. 2021). The range of the $^3$He/$^4$He ratios varies from 0.009 ± 0.001 to 41.3 ± 6.9, the latter event being observed by ACE on 2000 January 6 and whose calculated $^3$He/$^4$He agrees with that obtained by Mason et al. (2000). There is no indication that a $^3$He/$^4$He ratio of 41.3 is a physical limit. As expected, the median values are lower for ACE/ULEIS observations compared to that of STEREO/SIT due to its increased sensitivity toward $^3$He enhancements. Because of the superior mass resolution of ULEIS, the $^3$He/$^4$He (also Fe/O, see Figure 9) ratios extend much lower than those of SIT. For the same reason, ACE observations include events with small $^3$He-rich enhancements that are mixed with prior SEPs, CIRs, or IP shock events (see for example the 2011 February 18 $^3$He-rich SEP event in Bučík et al. 2018b). Mixing of SEPs from prior enhancements typically increases the abundance of $^4$He disproportionately to $^3$He. As a result, some of the low $^3$He/$^4$He ratios are considered to be lower limits, though the real $^3$He enrichment can be much larger.

The statistical properties listed in Figure 8, though themselves not indicative of the true distribution of $^3$He/$^4$He ratios in $^3$He-rich time periods, do provide some insight into the subject. As the mass resolution of the detector increases, the median $^3$He/$^4$He value decreases, highlighting that our sampled distribution is biased toward large $^3$He enhancements because they are easily visible in the four-day-plots. Indeed, this selection method excludes many large SEP events with only slightly enhanced $^3$He/$^4$He above solar wind values (see Desai et al. 2006), and it excludes events that have all the signatures of ISEP events (low scattering, type III burst, enhanced Fe/O) but no clear $^3$He/$^4$He enhancement. Overall, the statistical means and medians listed in Figure 8 are likely overestimates of the true $^3$He/$^4$He distributions during $^3$He-rich time periods.

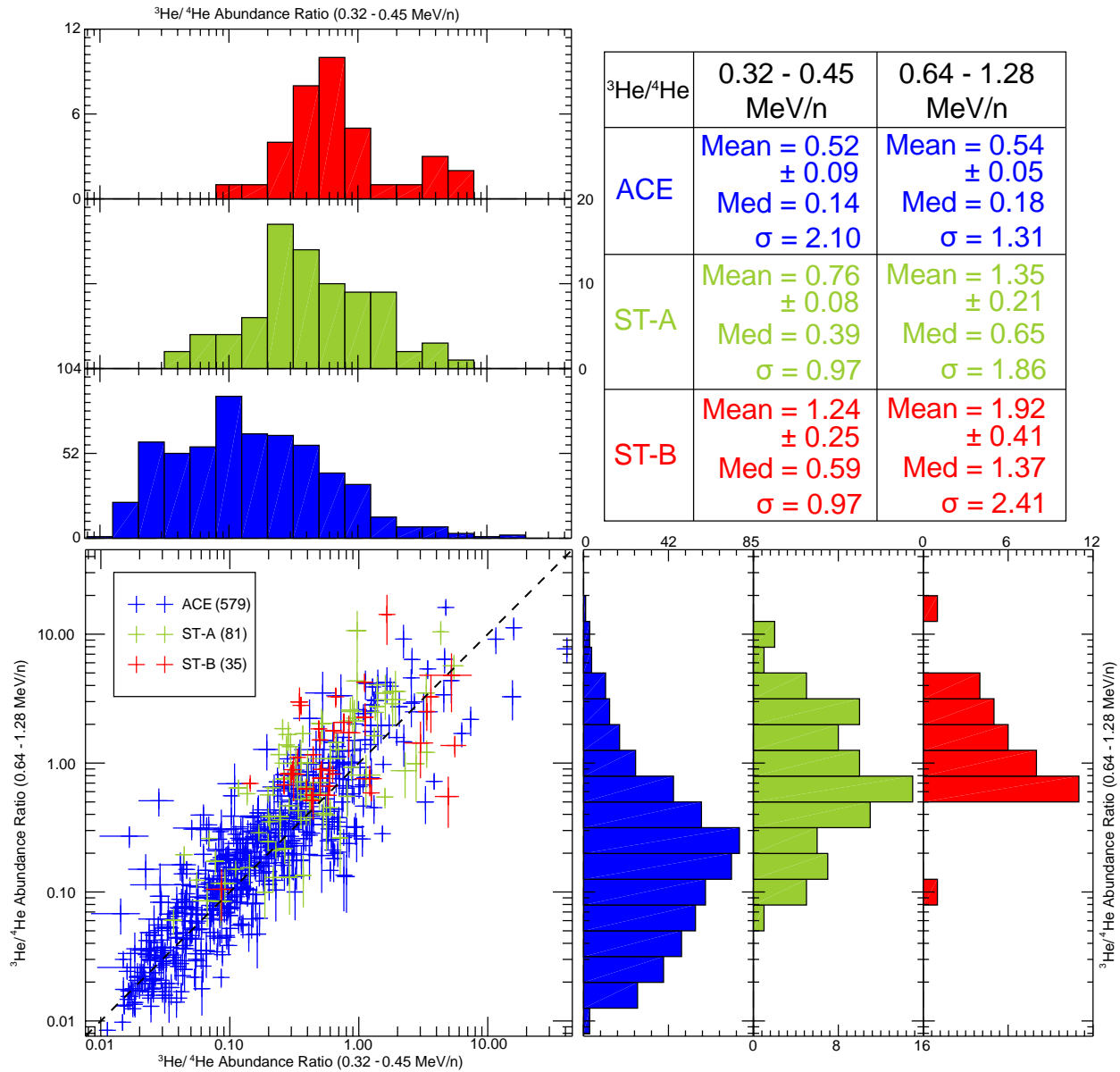

**Figure 8.** $^3$He/$^4$He ratios at 0.64 – 1.28 MeV/nucleon vs. 0.32 – 0.45 MeV/nucleon with a dashed line of symmetry. The 1-D ratio distributions for each spacecraft are shown along their respective axis, matching colors accordingly. The mean, median, and standard deviation of $^3$He/$^4$He for each spacecraft are shown in the top right of the figure.

Similar to Figure 8, Figure 9 shows the Fe/O abundance ratios at high energies vs. low energies. Just like $^3$He/$^4$He, the Fe/O ratios show a clear correlation between the two energy ranges. Fe/O displays a narrower spread ranging from 0.03 to 12.1. Furthermore, there is no consistent Fe/O ratio increase from the low to high energy range indicating that Fe and O typically have similar spectral shapes. We have added and labelled the nominal Fe/O abundance ratios found in ISEPs, GSEPs, CIRs, and the slow solar wind as vertical and horizontal lines (from Desai et al. 2006). Note that some ULEIS observations dip into these regimes in part due to event mixing described in the previous paragraphs as well as observations of $^3$He enhancements in event types other than ISEP events. However, both STEREO spacecraft cannot as easily resolve $^3$He other than in ISEP events and thus their Fe/O ratios lie nearly entirely between 1.0 and 3.0. Unlike $^3$He/$^4$He, the mode of the Fe/O distributions remain consistent across all three spacecraft. Thus, we believe the Fe/O distribution observed by ACE is an approximate representation of the true Fe/O distribution of $^3$He-rich time periods even though our selection criteria is not fully inclusive.

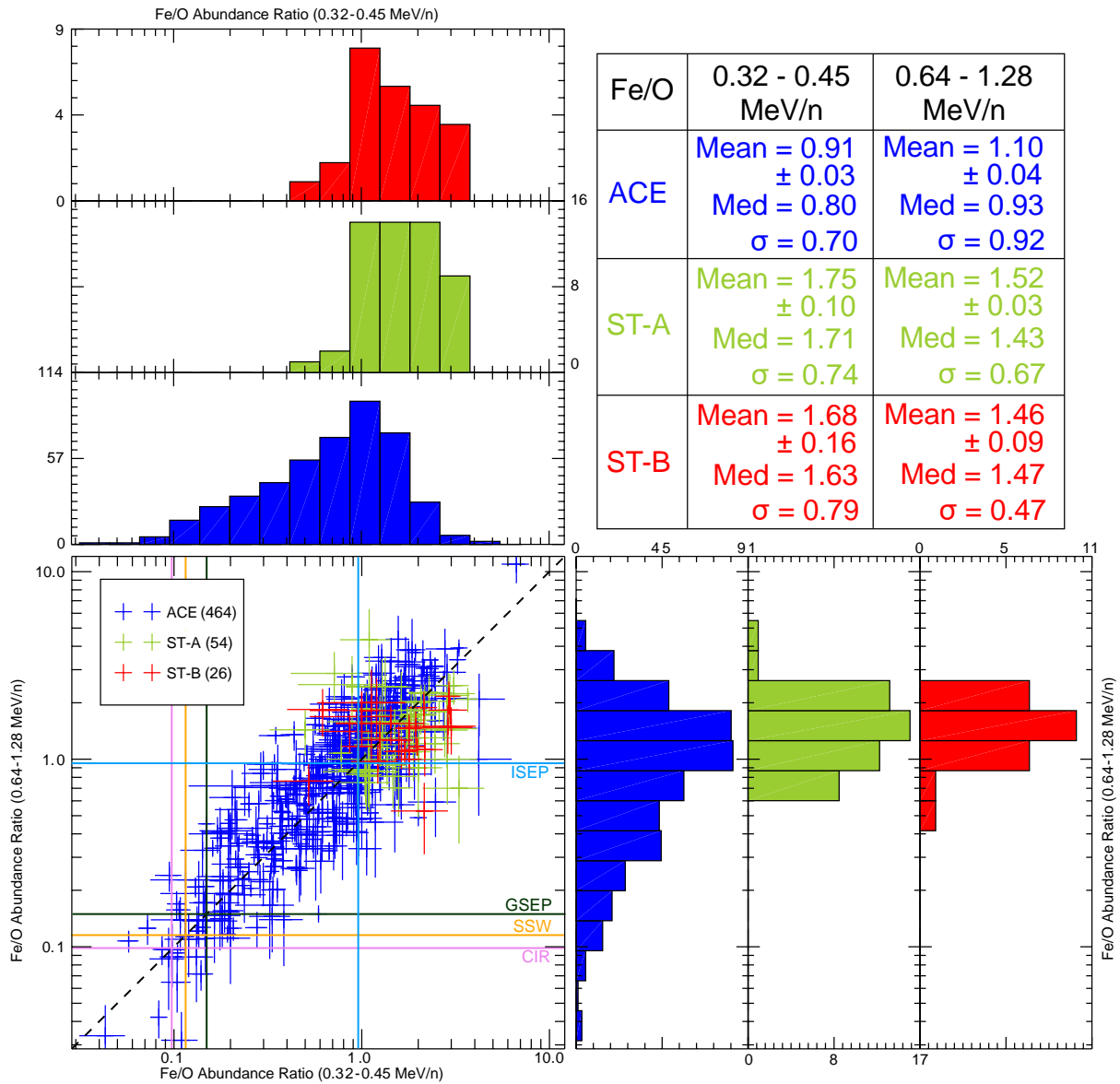

**Figure 9.** Fe/O abundances ratios at 0.64 – 1.28 MeV/nucleon vs. 0.32 – 0.45 MeV/nucleon with line of symmetry. Colored vertical & horizontal lines show the nominal Fe/O ratios of commonly observed heliospheric populations as a comparison. Again, the mean, median, and standard deviation of Fe/O for each spacecraft are shown in the top right of the figure, noting that (1) The median Fe/O ratios are a factor of 10 higher than GSEPs, CIRs, and the slow solar wind and (2) the standard deviation of the Fe/O ratios is substantially smaller than that of $^3$He/$^4$He.

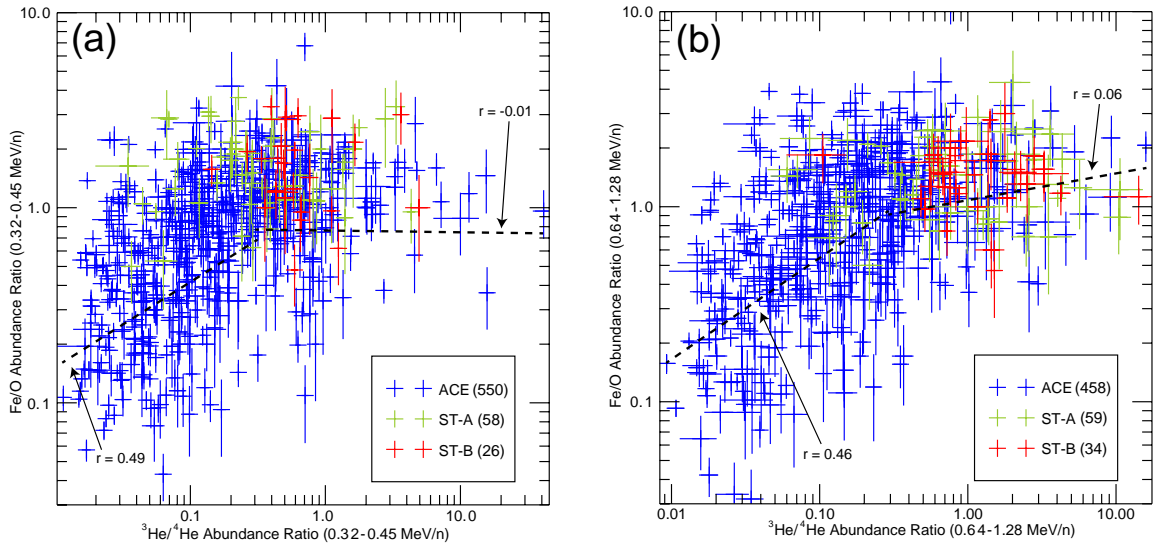

**Figure 10.** $^3$He/$^4$He vs. Fe/O for 0.32 – 0.45 MeV/nucleon (a) and 0.64 – 1.28 MeV/nucleon (b). At low $^3$He/$^4$He ratios, there is a moderate correlation with the Fe/O values ((a), r = 0.49; (b), r = 0.46). Above $^3$He/$^4$He = 0.3, the correlation tends to zero ((a), r = -0.01; (b), r = 0.06).

Figure 10 shows the Fe/O abundance ratios vs. the $^3$He/$^4$He abundance ratios for both energy ranges. We see two trends stemming from $^3$He/$^4$He ~ 0.3. Above this ratio, we obtain a correlation coefficient of -0.01 (a) and 0.06 (b) indicating no correlation between the two ratios in accordance with previous publications (Mason et al. 1986, Reames et al. 1994). Below this ratio, we obtain a positive correlation coefficient of 0.49 (a) and 0.46 (b). To fully understand this plot, we must consider all of the factors that influence our measured abundance ratios. This low-ratio correlation could be arising due to SEP mixing or $^3$He enhanced events other than ISEP events. However, it could also be the case that some ISEP events with little $^3$He enhancement are Fe poor as well. The latter case implies a dependence on the $^3$He/$^4$He enhancement up to a certain threshold where the Fe/O abundance ratio reaches a limit. Above this limit, Fe/O no longer increases as $^3$He/$^4$He increases resulting in the small correlation coefficients at large $^3$He/$^4$He ratios. This effect is consistent with the work reported by Dwyer et al. (2001).

## 5. Conclusion

This work serves to introduce a live catalogue of $^3$He-rich time periods using multiple spacecraft and covering multiple solar cycles. The release of this catalogue is timely with the recent launches of two new missions active in the inner heliosphere that are capable of observing $^3$He enhancements, namely, Parker Solar Probe ([Wiedenbeck et al. 2020](#)) and Solar Orbiter ([Mason et al. 2021](#), [Bučík et al. 2021](#)). The catalogue will be continuously updated, and major updates will be properly documented and reported in separate publications. Such updates may include adding data from new missions or deriving new parameters of the identified events. Remote sensing observations are also critical to better understand the underlying processes dominating $^3$He enhancements in the heliosphere. Future updates will include identifying the event types producing each enhancement and identifying the source locations within the active regions that produced these events using ion velocity dispersion back-tracking and associated remote observations from SOHO, SDO, STEREO, and radio observations from WIND and STEREO. When accompanying electrons are available, we will compare the path length and release times to the ions to better understand the underlying acceleration mechanisms (see [Li et al. 2020](#)).

We have identified 854 $^3$He-rich time periods observed by ACE and STEREO-A & B at 1 AU between 1997 September. – 2021 March . For each event and whenever possible, we determine the start and stop times of the event and determine the $^3$He/$^4$He and Fe/O abundance ratios. We have established two different techniques to separate $^3$He and $^4$He: a simple mass cutoff for ULEIS and heavy-ions, and a triple-peaked Gaussian fitter for SIT. Using these techniques, we have found that:

(1) $^3$He/$^4$He during $^3$He-rich time periods typically lie between 0.01 and 10, with a median value of 0.14, 0.39, 0.59 (ULEIS, SIT-A, SIT-B) for 0.32 – 0.45 MeV/nucleon and 0.18, 0.65, 1.37 for 0.64 – 1.28 MeV/nucleon. There is a strong correlation between the $^3$He/$^4$He abundance ratio at low energies and high energies.

(2) Fe/O during $^3$He-rich time periods typically lie between 0.1 and 5 with a median value of 0.80, 1.71, 1.63 (ULEIS, SIT-A, SIT-B) for 0.32 – 0.45 MeV/nucleon and 0.93, 1.43, 1.47 for 0.64 – 1.28 MeV/nucleon. The Fe/O abundance ratio also shows a strong correlation between low and high energies with a tighter spread when compared to $^3$He/$^4$He.

(3) The $^3$He/$^4$He vs. Fe/O abundance ratios show multiple trends. At larger $^3$He/$^4$He ratios, there is no correlation between Fe/O and $^3$He/$^4$He ratios. At smaller $^3$He/$^4$He ratios, we observe a moderate correlation between the two measurements. Beyond the scope of the initial catalogue release, further work is required to determine whether the observed low-ratio correlation is merely due to SEP mixing from preceding SEP enhancements and observations of $^3$He enhancements in GSEPs, shocks, and CIRs, or if smaller $^3$He/$^4$He enhancements in ISEP events are also accompanied by smaller Fe/O enhancements. This study requires each $^3$He-rich time period to be categorized into its respective event type. This can be done by referring to published event lists covering solar cycles 23 and 24. The Fe/O vs. $^3$He/$^4$He relation can then be analyzed for each event classification individually to determine which event type(s) maintains the trend seen in Figure 10.

## 6. Acknowledgements

The primary author would like to thank K. Moreland for useful discussions on the topic as well as those who have attended presentations of this work at various conferences. This work is currently supported by NASA's FINESST grant 80NSSC21K1389 and was previously supported by NASA's LWS grants 80NSSC19K0079 and 80NSSC21K1316.